\documentclass[pdftex]{elsart}

\usepackage{graphicx,color,xspace,amsmath,amssymb}
\providecommand{\MM}[1]{\left[ #1\right]}
\providecommand{\MMT}[1]{\left[ #1\right]^T}
\providecommand{\VV}[1]{\left\{ #1\right\}}
\providecommand{\PP}[1]{\left( #1\right)}
\providecommand{\dint}{\displaystyle \int }%
\providecommand{\cL}{\ensuremath{\mathcal{L}}\xspace}%

\renewcommand{\VV}[1]{\mathbf{\boldsymbol{#1}}}

\renewcommand{\MM}[1]{\mathbf{\boldsymbol{#1}}}
\renewcommand{\MMT}[1]{{}^t\!\mathbf{\boldsymbol{#1}}}
\newcommand{\MMI}[1]{\mathbf{\boldsymbol{#1}}^{-1}}
\newcommand{\ansys}{\textsc{Ansys}\xspace}

\definecolor{grisi}{gray}{0.25}
\definecolor{grisii}{gray}{0.5}
\definecolor{grisiii}{gray}{0.75}
\definecolor{grisiv}{gray}{0.5}
\definecolor{grisv}{gray}{0.5}

\begin{document}

\begin{frontmatter}
\title{%
  The use of \ansys to calculate sandwich structures
}
\author{%
   Vincent \textsc{Manet}
}
\address{%
      \'Ecole des Mines de Saint-Etienne,
      Material and Mechanical Department,
      158, cours Fauriel,
      42023 Saint-Etienne cedex 2,
      France\\
      fax: (+33) 4-77-42-00-00,
      email: \texttt{manet@emse.fr}
}


\begin{abstract}
   In this article, we make a comparative study on a simply
   supported sandwich beam subjected to a uniform pressure
   using different modellings offered by the software \ansys 5.3
   to compute displacements and stresses.

   8 nodes quadrilateral elements Plane 82, multi-layered 8 nodes
   quadrilateral shell element Shell 91 and multi-layered 20 nodes
   cubic element Solid 46 are used.
   Influence of mesh refinement and of ratio of young's moduli of
   layers are studied.

   Finally, a local Reissner method is presented and assessed, which
   permits to improve the accuracy of interface stresses for high ratio
   of young's moduli of layers using Plane 82 elements.
\end{abstract}

\end{frontmatter}

\medskip
\section{Introduction}

Sandwich materials really begin to be highly appreciated in
the industry, and especially in the field of transport (automotive,
aeronautics, shipbuilding and railroads) or in civil engineering.

\medskip
It is therefore important to determine which elements should be
used to model such structures.

\medskip
A sandwich structure is composed of three layers:
\begin{itemize}
   \item two edges made of rigid layers, working in membrane,
         which represent the skins;
   \item a thick and soft central layer, the core, with low
         rigidity and density and essentially submitted to transverse
         shear loading, is sandwiched in between the edges.
\end{itemize}

\medskip
In the design process, interface stresses can be of great importance,
since they play a crucial role in failure modes, as explained in
\cite{Teti-Caprino,couvrat}.

\medskip
The core being essentially subjected to transverse shear stress,
this component, which is generally very much lower than the others,
must not be neglected: in some cases, effects arising from
shear effects overhang others phenomena (flexural effects for
example), as shown for example in \cite{Zenkert,Allen,allen,hoff}.

\medskip
The determination of transverse shear stress at interfaces is therefore
of particular importance in the design of new optimized materials.

\medskip
If we assume that the three layers remain perfectly bonded, then
at interfaces:
\begin{itemize}
   \item the displacement field must be continuous;
   \item the normal trace of the stress must be continuous.
\end{itemize}

\medskip
In this article, we shall study a very simple case using the famous
finite element software \ansys 5.3.
We shall not talk about special elements based on
hybrid \cite{csma97,euromech97},
mixed \cite{Aiva1,Aiva2,bichara,lardeur}
or modified \cite{verchery,phamdang,elshaikh}
formulation nor about pre- and post-processing methods \cite{Pai,lerooy}.

\medskip
Solutions obtained with different modellings (complex or simple ones)
are compared. Particular emphasis is put on their respect of continuity
requirements.

By modifying the stiffness of the core, we shall see which modelling
should be preferred by designers.

Finally a method, based on Reissner's formulation, is developed
to improve the accuracy for new sandwiches.

\medskip
\section{Description of the study}

\begin{figure}[ht]
   \begin{center}
      \includegraphics[height=\textwidth,angle=-90]{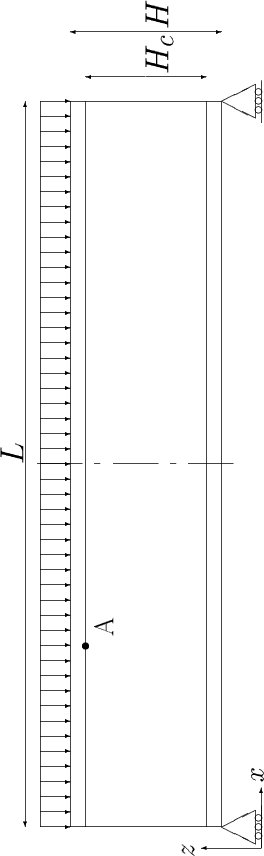}
   \end{center}
   \caption{\label{Fig:def-poutre} Sandwich beam}
\end{figure}

One of the most simple example is the case of the simply supported
sandwich beam subjected to an uniform pressure on its top face.
This beam is shown in figure \ref{Fig:def-poutre}.

\medskip
\subsection{Characteristics}

\medskip
The geometry is defined as follows:
\begin{itemize}
   \item The total length of the beam is $L=24$~mm;
   \item its total height $H=2$~mm, the core representing
         80\% of the total height of this symmetrical
         sandwich, each skin is $0.2$~mm height;
   \item the thickness of the beam is equal to unity.
\end{itemize}

The applied pressure is $q=-1$~N/mm.

\medskip
\subsection{Parameters of the study}

In this study, we are interested in determining the structural
response at point A (at the interface between the top skin and
the core and located at $x_1=L/4$)
when different parameters vary.

\medskip
Skins are made of aluminum ($E_s=70000$~MPa and $\nu_p=0.34$).
The core will be:
\begin{itemize}
   \item Case A: of carbon/epoxy ($E_c=3400$~Mpa and $\nu_c=0.34$);
   \item Case B: of foam ($E_c=340$~Mpa and $\nu_c=0.40$);
   \item Case C: of soft foam ($E_c=70$~Mpa and $\nu_c=0.40$);
   \item Case D: other material: $\nu_c=0.4$ is fixed and $E_s/E_c$ varies.
\end{itemize}

\medskip
\subsection{The modellings}

By symmetry, only one half of the beam is modelled.

\medskip
Before building the different modellings, we define the
following parameters for the meshing:
\begin{itemize}
   \item $ncuts$: number of longitudinal cuts (in the beam's axis
         direction);
   \item $nskin$: number of elements in the thickness of each skin;
   \item $ncore$: number of elements in the thickness of the core.
\end{itemize}

\medskip
We shall use the following modellings:
\begin{itemize}
   \item the reference modelling:
      \begin{itemize}
         \item 2D using the 8 nodes quadrilateral element Plane 82;
         \item 4 elements through the thickness of each skin ($nskin=4$),
               32 through the thickness of the core ($ncore=32$),
               and 400 longitudinal cuts ($ncuts=400$) in the beam's axis
               directions (16000 elements for the half beam);
      \end{itemize}
         This fine meshing yields to the exact solution given
         by \cite{pagano}.
   \item a planar modelling using the plane element Plane 82:

         1 element is used to model each layers (nskin = ncore = 1),
         i.e. 3 elements through the thickness of the sandwich;

   \item a modelling using the multi-layered cubic element Solid 46:

         1 element through the total thickness representing all
         the layers of the sandwich structure;

   \item one modelling done with the multi-layered shell element Shell 91,
         with sandwich option (keyopt(9)=1):

         1 element through the total thickness representing all
         the layers of the sandwich structure.
\end{itemize}

\medskip
\subsection{Results of interest}

In our studies, we shall focus on the following results of particular
interest:
\begin{itemize}
   \item the maximum displacement of the structure in $z$ direction, denoted
         $U_z$ in results;
   \item the discontinuous components of stresses, $\sigma_{xx}$, at
         point A in the skin and in the core, and the continuous
         component $\sigma_{zz}$;
   \item interlaminar stress: this is the continuous component $\sigma_{xz}$
         at point A.
\end{itemize}

\medskip
\section{Study of the sandwich beam}

We now present results obtained with \ansys 5.3 and corresponding
to different materials and different meshes.

\medskip
\subsection{Influence of \emph{ncuts} on the different modellings}
\label{Sec:ncuts}

In this section, we are interested in the structural responses to
the different modellings for $E_c=3400$~MPa, $E_c=340$~MPa and
$E_c=70$~MPa.

\medskip
Results concerning the case $E_c=340$~MPa are plotted in figures
\ref{Fig:uz0340} for displacements, \ref{Fig:sz0340} and
\ref{Fig:sxz0340} for the two continuous components $\sigma_{zz}$
and $\sigma_{xz}$ and \ref{Fig:sxc0340} and \ref{Fig:sxs0340}
for $\sigma_{xx}$ in the core and in the skin respectively.

\medskip
Tables \ref{Tab:3400}, \ref{Tab:0340} and \ref{Tab:0070} present
numerical results and error percentages after convergence
for these 3 cases.

\medskip
From these figures and tables, the following conclusions can be drawn:
\begin{itemize}
   \item Plane 82 is very much better than others modellings.
         Nevertheless, it is to be noticed that it seems to diverge
         for displacements (with the coarse mesh used: $nskin=ncore=1$);
   \item Solid 46 is the worst model. It never converges towards the
         right values (for any component of stresses nor for displacements);
   \item Shell 91 is particularly interesting for continuous
         components of stresses $\sigma_{zz}$ and $\sigma_{xz}$;
   \item Plane 82 is the only modelling leading to a correct determination
         of the discontinuous component $\sigma_{xx}$ in the skin and
         the core;
   \item It seems that errors increase with the ratio $E_s/E_c$. This
         point will be studied in the next section.
\end{itemize}

\medskip
\subsection{Influence of ratio $E_s/E_c$ for \emph{ncuts=20}}
\label{Sec:ratio}

Since every material which can be obtained in a thin skin shape
is acceptable for the skins and every material with low density
is acceptable for the core, sandwich materials cover an extremely
wide domain.

\medskip
A parameter of interest is therefore the ratio of young's moduli
$E_s/E_c$.
This parameter can vary from 4 (old sandwiches, so to speak, very
close to laminates) to 1000 (some new high-tech sandwiches developed
for very particular applications go up to 1500).
But we must remark that sandwiches often exhibit a ratio
greater than 200.

\medskip
In this section we shall study the influence of this ratio on the
different modellings when the utilized mesh is fixed to $ncuts=20$.

\begin{figure}[ht]
   \begin{center}
      \includegraphics{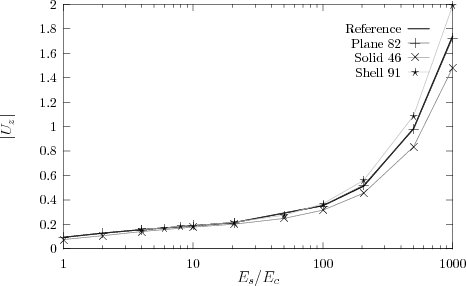}
   \end{center}
   \caption{\label{Fig:var-uz} Influence of $E_s/E_c$: $U_{z}$ ($ncuts=20$)}
\end{figure}

\begin{figure}[ht]
   \begin{center}
      \includegraphics{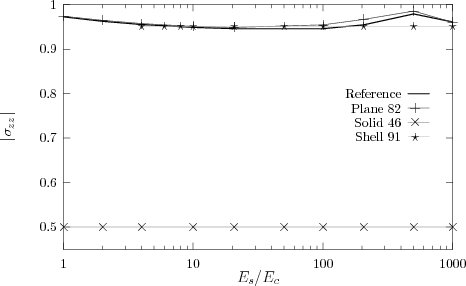}
   \end{center}
   \caption{\label{Fig:var-sz} Influence of $E_s/E_c$: $\sigma_{zz}$ ($ncuts=20$)}
\end{figure}

\begin{figure}[ht]
   \begin{center}
      \includegraphics{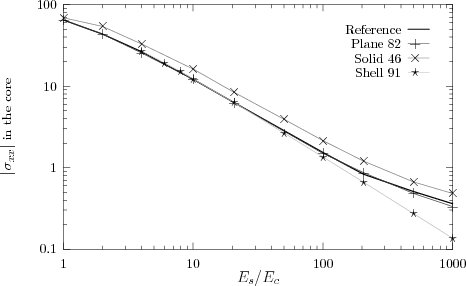}
   \end{center}
   \caption{\label{Fig:var-sxc} Influence of $E_s/E_c$: $\sigma_{xx}$ in the core ($ncuts=20$)}
\end{figure}

\begin{figure}[ht]
   \begin{center}
      \includegraphics{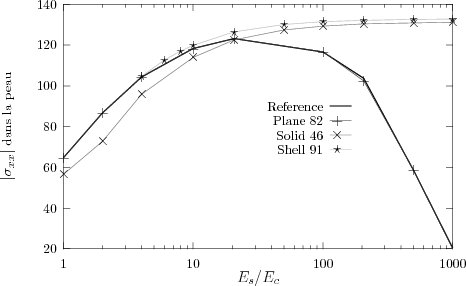}
   \end{center}
   \caption{\label{Fig:var-sxs} Influence of $E_s/E_c$: $\sigma_{xx}$ in the top skin ($ncuts=20$)}
\end{figure}

\medskip
Results concerning displacements are plotted in figure
\ref{Fig:var-uz}. Continuous components $\sigma_{zz}$ and $\sigma_{xz}$
are shown in figures \ref{Fig:var-sz} and \ref{Fig:var-sxz}.
The discontinuous component $\sigma_{xx}$ is illustrated in figures
\ref{Fig:var-sxc} and \ref{Fig:var-sxs} in the core and in the skin
respectively.

\medskip
From these figures, the following conclusions can be drawn:
\begin{itemize}
   \item Plane 82 is the best modelling for displacements, $\sigma_{zz}$
         and $\sigma_{xx}$ in the core and the skin;
   \item Shell 91 and Solid 46 are acceptable for displacements and
         $\sigma_{xx}$ in the core. They are acceptable for $\sigma_{xx}$
         in the skin for $E_s/E_c\le$ 50;
   \item Shell 91 leads to an acceptable approximation of $\sigma_{zz}$,
         and is very interesting for $\sigma_{xz}$;
   \item Plane 82, which was exceptionally good in the last section,
         shows some difficulty here, especially at high $E_s/E_c$ ratio
         for $\sigma_{xz}$.
         The influence of the meshing of the beam with Plane 82
         elements is studied in the next section.
\end{itemize}

\medskip
\subsection{Element Plane 82: influence of mesh refinement}
\label{Sec:mesh}

In previous sections, the mesh corresponding to the 8 nodes
quadrilateral element Plane 82 only used 1 element to model
1 layer.

\medskip
We propose to see what happens when the number of elements through
the thickness of the skins ($nskin$) and of the core ($ncore$) vary.

\begin{figure}[ht]
   \begin{center}
      \includegraphics{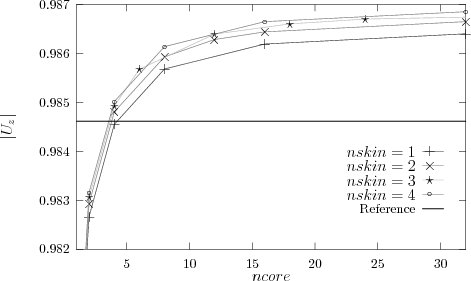}
   \end{center}
   \caption{\label{Fig:cv-uz} Influence of \emph{nskin} and \emph{ncore}: $U_{z}$ ($ncuts=20$ and $E_s/E_c=500$)}
\end{figure}

\begin{figure}[ht]
   \begin{center}
      \includegraphics{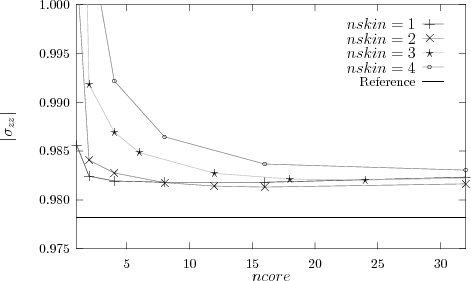}
   \end{center}
   \caption{\label{Fig:cv-sz} Influence of \emph{nskin} and \emph{ncore}: $\sigma_{zz}$ ($ncuts=20$ and $E_s/E_c=500$)}
\end{figure}

\begin{figure}[ht]
   \begin{center}
      \includegraphics{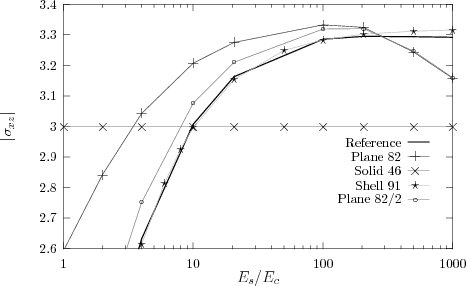}
   \end{center}
   \caption{\label{Fig:cv-sxz} Influence of \emph{nskin} and \emph{ncore}: $\sigma_{xz}$ ($ncuts=20$ and $E_s/E_c=500$)}
\end{figure}

\begin{figure}[ht]
   \begin{center}
      \includegraphics{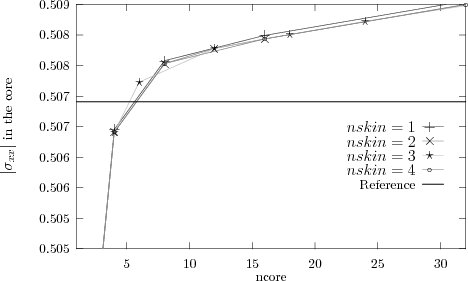}
   \end{center}
   \caption{\label{Fig:cv-sxc} Influence of \emph{nskin} and \emph{ncore}: $\sigma_{xx}$ in the core ($ncuts=20$ and $E_s/E_c=500$)}
\end{figure}

\begin{figure}[ht]
   \begin{center}
      \includegraphics{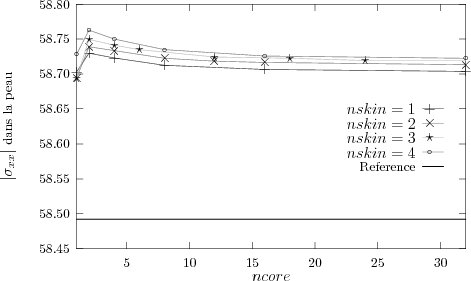}
   \end{center}
   \caption{\label{Fig:cv-sxs} Influence of \emph{nskin} and \emph{ncore}: $\sigma_{xx}$ in the top skin ($ncuts=20$ and $E_s/E_c=500$)}
\end{figure}

\medskip
Displacements are plotted in figure \ref{Fig:cv-uz},
$\sigma_{zz}$ and $\sigma_{xz}$ in figures \ref{Fig:cv-sz} and
\ref{Fig:cv-sxz}, and $\sigma_{xx}$ in figures \ref{Fig:cv-sxc}
and \ref{Fig:cv-sxs} in the core and in the skin respectively.
These computations are done for a ratio $E_s/E_c=500$.

\medskip
From these figures, the following conclusions can be drawn:
\begin{itemize}
   \item results are always accurate when $ncore = 8\  nskin$, i.e.
         when the meshing is regular through the thickness;
   \item $nskin$ and $ncore$ do not have any influence on the convergence
         of displacements, essentially due to flexion: the number
         of longitudinal cuts, $ncuts$, is therefore the most preponderant
         parameter.
   \item a very refined mesh ($nskin=4$ and $ncore=8\  nskin$)
         must be used in order to converge towards the right value
         of $\sigma_{zz}$;
   \item a coarse mesh ($nskin=1$) does not permit to obtain an acceptable
         value of $\sigma_{xz}$;
   \item convergence towards $\sigma_{xx}$ reference value in the core
         is controlled by ncuts. Results are not improved by increasing
         $nskin$ nor $ncore$;
   \item the last point is also true for the convergence towards
         $\sigma_{xx}$ value in the skin.
\end{itemize}

\medskip
\section{Local Reissner: improving results for Plane 82}

\begin{figure}[ht]
   \begin{center}
      \includegraphics{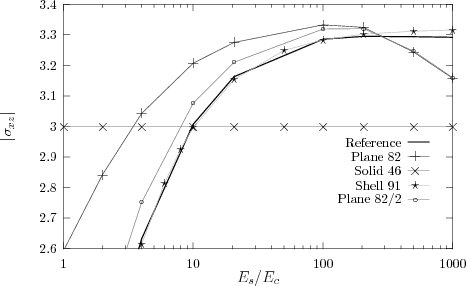}
   \end{center}
   \caption{\label{Fig:var-sxz} Influence of $E_s/E_c$: $\sigma_{xz}$ ($ncuts=20$)}
\end{figure}

As it can be see from figure \ref{Fig:var-sxz} and from table
\ref{Tab:accu} (which summaries results and gives the good ``working
zone'' of the different modellings), Plane 82 is not able to
give accurately the interlaminar stress $\sigma_{xz}$ with
a coarse mesh.
Since this component is very important in the design process,
results must be improved.

\medskip
A way of improving results is to refine the meshing.
In figure \ref{Fig:var-sxz}, the curve `Plane 82/2' gives results
obtained with $nskin = 1$ and $ncore = 2$ (instead of 1).
This slightest modification of the mesh (4 elements through the
thickness of the sandwich instead of 3) is sufficient to lead to
very good results for $E_s/E_c\le200$.

\medskip
But, as mentioned before, sandwiches nowadays exhibit ratios generally
higher than 200.
In this range, the convergence is only reached with a very refined meshing:
$nskin\ge 3$ and $ncore = 8\  nskin$.
Such a mesh yields an unacceptable computation time.

\medskip
In order to improve the accuracy of stresses, we must answer to the
following question: how are nodal stresses computed?

\medskip
Nodal stresses $\VV{\tau}$ are generally computed using a
minimization process. They are obtained from nodal displacements
$\VV{q}$ using a least squares method and by minimizing:
\begin{equation}
   \label{Eq:carre}
   \dint_\Omega \PP{\VV{\sigma_m}-\VV{\sigma_u}}^2\ \d\Omega
\end{equation}
where $\VV{\sigma_m}$ denotes the mixed way to calculate stresses:
\begin{equation}
   \label{Eq:ProjSigmaMixte}
   \VV{\sigma_m} = \MM{N_\sigma}\VV{\tau}
\end{equation}
and $\VV{\sigma_u}$ the displacements way:
\begin{equation}
   \label{Eq:ProjSigmaDisp}
   \VV{\sigma_u} = \MM{D}\MM{\cL}\MM{N_u}\VV{q}
\end{equation}
or using the stress projection method presented in \cite{zien} by
minimizing:
\begin{equation}
   \label{Eq:strproj}
   \dint_\Omega \PP{\VV{\sigma_m}-\VV{\sigma_u}}\ \d\Omega
\end{equation}
In these equations, $\MM{D}$ is the generalized Hooke's matrix relating
stresses to strains, $\MM{\cL}$ the differential operator relating strains
to displacements, $\MM{N_\sigma}$ and $\MM{N_u}$ the matrices of shapes
functions for stresses and displacements, and $\Omega$ the volume or
surface of interest.

\medskip
It is to be noticed that these methods lead to convergence towards
Reissner's (reference) solution.

\medskip
As expressed in \cite{hinton}, the minimization process can be
global (done over the whole structure: $\Omega$ being the entire
structure) or local (done over one element: $\Omega$ being the
considered element).
Since the local process converges towards the same limit as the
global process, the minimization process chosen is generally the
local one.

\medskip
Nevertheless, instead of minimizing the difference between two
solutions, it may be more convenient (simpler and faster) to
directly find the stress field using Reissner's formulation.

\medskip
In Reissner's solution, nodal stresses are related to nodal
displacements by \cite{Reiss0,Reiss1,Reiss2,Washizu}:
\begin{equation}
   \label{Eq:tauq}
   \VV{\tau}=\MMI{A}\MM{B}\VV{q}
\end{equation}
with:
\begin{equation}
   \label{Eq:A}
   \MM{A}=+\dint_\Omega \MMT{N_\sigma}\MM{S}\MM{N_\sigma}\ d\Omega
\end{equation}
and
\begin{equation}
   \label{Eq:B}
   \MM{B}=+\dint_\Omega \MMT{N_\sigma}\MM{\cL}\MM{N_u}\ d\Omega
\end{equation}
$\MM{S}=\MMI{D}$ being the compliance matrix.

\medskip
In order to improve the stress computation at interfaces, we propose
to use the last formulation on \emph{two} adjacent elements, located
on each side of an interface. Doing so, we ensure the equilibrium
state at interface on a better way.
We shall refer to this method as ``local Reissner'' method.

\medskip
This kind of method is not more time consuming than least squares
methods generally used (in \ansys for example) to derive nodal
stresses from nodal displacements.

\begin{figure}[ht]
   \begin{center}
      \includegraphics{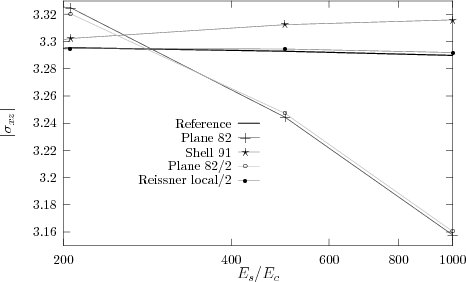}
   \end{center}
   \caption{\label{Fig:var-sxz-} Influence of high $E_s/E_c$ ratio: $\sigma_{xz}$ ($ncuts=20$)}
\end{figure}

\medskip
Now looking at figure \ref{Fig:var-sxz-}, which is a close-up view
of figure \ref{Fig:var-sxz} for sandwiches with $E_s/E_c\ge200$,
we can see that:
\begin{itemize}
   \item the use of local Reissner's method (denoted Local Reissner/2
         because we used $ncore = 2$) permits to really improve the
         accuracy of $\sigma_{xz}$, which is of great importance.
   \item It is to be noticed that the same mesh with Plane 82 (Plane
         82/2) does not permit to improve results with this high
         $E_s/E_c$ ratio.
   \item The solution given by Shell 91 is not so good as for lower
         $E_s/E_c$ ratio.
\end{itemize}

\medskip
\section{Conclusions}

The reference solution has been obtained using a very fine meshing
and the 8 node quadrilateral element Plane 82 in order to reach
the analytical solution of Pagano \cite{pagano}.

\begin{figure}[ht]
   \begin{center}
      \includegraphics{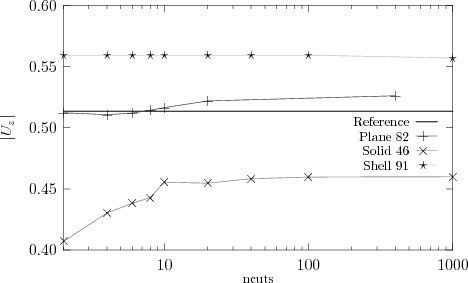}
   \end{center}
   \caption{\label{Fig:uz0340} Influence of \emph{ncuts}: $U_{z}$ (case B)}
\end{figure}

\begin{figure}[ht]
   \begin{center}
      \includegraphics{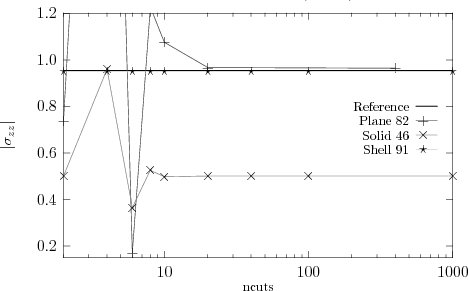}
   \end{center}
   \caption{\label{Fig:sz0340} Influence of \emph{ncuts}: $\sigma_{zz}$ (case B)}
\end{figure}

\begin{figure}[ht]
   \begin{center}
      \includegraphics{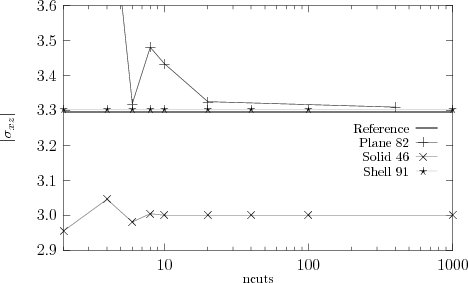}
   \end{center}
   \caption{\label{Fig:sxz0340} Influence of \emph{ncuts}: $\sigma_{xz}$ (case B)}
\end{figure}

\begin{figure}[ht]
   \begin{center}
      \includegraphics{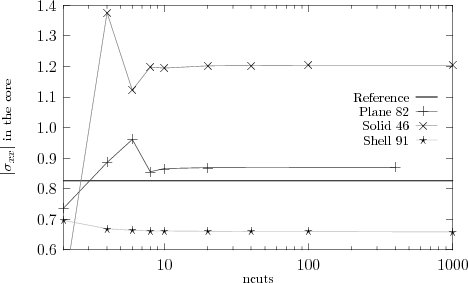}
   \end{center}
   \caption{\label{Fig:sxc0340} Influence of \emph{ncuts}: $\sigma_{xx}$ in the core (case B)}
\end{figure}

\begin{figure}[ht]
   \begin{center}
      \includegraphics{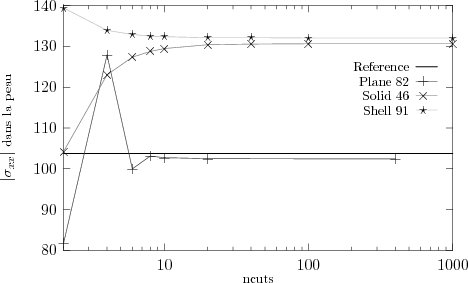}
   \end{center}
   \caption{\label{Fig:sxs0340} Influence of \emph{ncuts}: $\sigma_{xx}$ in the top skin (case B)}
\end{figure}

\medskip
The first study, influence of \emph{ncuts} on the different modelling
in section \ref{Sec:ncuts}, seems to lead to the conclusion that
Plane 82 is the best model, specially when looking at figures
\ref{Fig:uz0340}-\ref{Fig:sxs0340} (obtained with a coarse mesh
and in the case $E_s/E_c\approx200$).
Nevertheless, the study of the influence of the $E_s/E_c$ ratio
in section \ref{Sec:ratio} permits to see some weakness of this
model.

\medskip
In terms for design quantities:
\begin{itemize}
   \item all models leads to a correct value of displacements, but
         Plane 82 is the most accurate;
   \item $\sigma_{zz}$ can be correctly given by Shell 91 and Plane 82,
         the latest being the most accurate;
   \item $\sigma_{xz}$ is only very accurately computed with Shell 91, but
         for $E_s/E_c\le 200$;
   \item $\sigma_{xx}$ in the core can be calculated using any model,
         Plane 82 being the most accurate;
   \item $\sigma_{xx}$ in the skin is very accurately computed with
         Plane 82 and with Shell 91 (but only for $E_s/E_c\le 20$), and
         acceptable with Solid 46 (and only for $E_s/E_c\le 20$).
\end{itemize}

\medskip
A summary of results, and the ``working zone'' in which the different
elements can be used is given in table \ref{Tab:accu}.

\medskip
Hence, from the previous results, we can say that:
\begin{itemize}
   \item for planar problems, Plane 82 is very well adapted. Nevertheless,
         it is to be noticed that this method is not very stable for
         very coarse meshes (small values of $nskin$, $ncore$ and
         $ncuts$), and that interlaminar stress $\sigma_{xz}$ can
         only be reached with a fine meshing;
   \item Shell 91 (with sandwich option) is a good way of computing
         sandwich structures.
         Nevertheless if the designer must know $\sigma_{xx}$ at
         interfaces, then this element can only be used for
         $E_s/E_c\le50$.
   \item Solid 46 is not very accurate in the determination of
         the design quantities. A model using this kind of element
         should be avoided. Nevertheless, it is to be noticed that
         this element has not been developed to perform such
         computations (high differentce of stiffness between
         layers).
\end{itemize}

\medskip
The presented local Reissner's method permits to reach excellent
results, especially for the interlaminar stress $\sigma_{xz}$
and for $E_s/E_c\ge200$ with a coarse mesh through the thickness
of the sandwih. We can notice that such a meshing, with 4 elements
through the thickness of the sandwich yields results very closed
to the exact solution obtained with 40 elements through the
thickness!

This method is particularly interesting for the design of new
sandwich materials.

\medskip
Finally, we want to put emphasis on the fact that this method is
particularly easy to implement, as a stand-alone program, but also
in existing finite element softwares.

\renewcommand{\baselinestretch}{1}

\newpage
\cleardoublepage

\cleardoublepage

\bigskip
~
\bigskip

\cleardoublepage

\newpage
\renewcommand{\baselinestretch}{1}

\begin{table}[ht]
   \caption{\label{Tab:3400} Case A: $E_c=3400$~Mpa}
   \begin{center}
      \begin{tabular}{rrrrrrr}
         \multicolumn{7}{l}{}\\
         \hline
         & $|U_z|$ & $|\sigma_{zz}|$ &$|\sigma_{xz}|$ & $|\sigma_{xx}|$ & $|\sigma_{xx}|$ & ncuts\\[-3mm]
         & & & & skin & core & \\
      \hline
      Ref& 0.21596 & 0.94624 & 3.1635 & 123.13 & 6.2844 &\\
      P82& 0.21527 & 0.95033 & 3.2616 & 123.08 & 6.2843 & 400\\
         & 0.319\% & 0.432\% & 3.10\% & 0.04\% & 0.001\%& \\
      S91& 0.21388 & 0.90000 & 3.1587 & 126.35 & 6.1369 & 1000\\
         & 0.963\% & 4.887\% & 0.15\% & 2.61\% & 2.35\% &\\
      S46& 0.20355 & 0.50000 & 3.0000 & 123.28 & 8.4202 & 1000\\
         & 5.746\% & 47.16\% & 5.17\% & 0.12\% & 34.0\% &\\
      \hline
      \end{tabular}
   \end{center}
\end{table}

\begin{table}[ht]
   \caption{\label{Tab:0340} Case B: $E_c=340$~Mpa}
   \begin{center}
      \begin{tabular}{rrrrrrr}
         \multicolumn{7}{l}{}\\
         \hline
         & $|U_z|$ & $|\sigma_{zz}|$ &$|\sigma_{xz}|$ & $|\sigma_{xx}|$ & $|\sigma_{xx}|$ & ncuts\\[-3mm]
         & & & & skin & core & \\
      \hline
      Ref& 0.51353 & 0.95461 & 3.2956 & 103.71 & 0.8262&\\
      P82& 0.52618 & 0.96321 & 3.3091 & 102.38 & 0.8701& 400\\
         & 2.463\% & 0.901\% & 0.41\% & 1.28\% & 5.31\%&\\
      S91& 0.55712 & 0.90000 & 3.3024 & 132.09 & 0.6599& 1000\\
         & 8.488\% & 5.721\% & 0.22\% & 27.4\% & 20.1\%&\\
      S46& 0.45994 & 0.50000 & 3.0000 & 130.68 & 1.2042& 1000\\
         & 10.44\% & 47.62\% & 8.97\% & 26.0\% & 45.7\%&\\
      \hline
      \end{tabular}
   \end{center}
\end{table}

\begin{table}[ht]
   \caption{\label{Tab:0070} Case C: $E_c=70$~Mpa}
   \begin{center}
      \begin{tabular}{rrrrrrr}
         \multicolumn{7}{l}{}\\
         \hline
         & $|U_z|$ & $|\sigma_{zz}|$ &$|\sigma_{xz}|$ & $|\sigma_{xx}|$ & $|\sigma_{xx}|$ & ncuts\\[-3mm]
         & & & & skin & core & \\
      \hline
      Ref& 1.740  & 0.96121 & 3.2926 &  20.43 & 0.3614&\\
      P82& 1.741  & 0.96847 & 3.1561 &  20.77 & 0.3301& 400\\
         & 0.06\% & 0.755\% & 4.15\% & 1.66\% & 8.66\%&\\
      S91& 1.987  & 0.90000 & 3.3161 & 132.64 & 0.1364&1000\\
         & 14.2\% & 6.368\% & 0.71\% & 549 \% & 62.2\%&\\
      S46& 1.506  & 0.50000 & 3.0000 & 131.43 & 0.4802&1000\\
         & 13.4\% & 47.98\% & 8.89\% & 543 \% & 32.9\%&\\
      \hline
      \end{tabular}
   \end{center}
\end{table}


\begin{table}[ht]
   \caption{\label{Tab:accu} Accuracy of results (ncuts $\ge 20$ understood)}

   \begin{center}
      \begin{tabular}{r|p{22mm}|p{22mm}|p{22mm}|p{22mm}|p{22mm}}

          \multicolumn{6}{l}{}\\
          \hline
             & $U_z$
             & $\sigma_{zz}$
             & $\sigma_{xz}$
             & $\sigma_{xx}$
             & $\sigma_{xx}$\\[-2.5mm]
             &&&& core & skin\\
         \hline
         P82
             & always good 
             & always good 
             & acceptable for $E_s/E_c \in [100,400]$ 
             & always good 
             & always good \\
             & \multicolumn{5}{l}{improvement of results}\\
             & \multicolumn{5}{l}{-- use a fine meshing:
               $nskin \ge 3$ and $ncore = 8\  nskin$}\\
             & \multicolumn{5}{l}{-- use local Reissner's method}\\
         \hline
         S91
             & always acceptable
             & always acceptable
             & always good
             & good for $E_s/E_c \le 100$
             & good for $E_s/E_c \le 20$\\
         \hline
         S46
             & always acceptable
             & never acceptable
             & acceptable for $E_s/E_c \in [8,15]$ 
             & always weak
             & acceptable for $E_s/E_c \le 20$\\
         \hline
      \end{tabular}
   \end{center}
\end{table}


\end{document}